\documentclass[10pt,conference]{IEEEtran}

\begin{document}

\title{Pattern Recognition System Design with Linear Encoding for Discrete Patterns}

\author{\authorblockN{Po-Hsiang Lai}
\authorblockA{Department of Electrical and Systems Engineering\\
Washington University in Saint Louis\\
Saint Louis, Missouri, 63130  USA \\
Email: pl1@cec.wustl.edu}
\and
\authorblockN{Joseph A.~O'Sullivan}
\authorblockA{Department of Electrical and Systems Engineering\\
Washington University in Saint Louis\\
Saint Louis, Missouri, 63130  USA \\
Email: jao@wustl.edu}
 }
%

\maketitle

\begin{abstract}
Pattern recognition systems based on compressed patterns and compressed sensor measurements can be designed using low-density matrices.  We examine truncation encoding where a subset of the patterns and measurements are stored perfrectly while the rest is discarded.  We also examine the use of LDPC parity check matrices for compressing measurements and patterns.  We show how more general ensembles of good linear codes can be used as the basis for pattern recognition system design, yielding system design strategies for more general noise models.
\end{abstract}



\section{INTRODUCTION}
\label{sect:intro}  
A recognition system has to be able to survive in a noisy environment subject to its own resource constraints. In most cases, including animals and machines, memory sizes are finite and sensory systems are only capable of extracting a fraction of information about an existing object. Also, a network decision system may consist of a sensing agent at one location, a database at a second location, and an action agent at a third location.  The action agent needs data from the sensor and the database for recognition and subsequent actions. With bandwidth limitations on communication channels, the action agent must perform recognition based on compressed, maybe lossy, data from the sensing agent and the database. Westover \cite{Westover06} and Westover and O'Sullivan \cite{Westover06b} derive inner and outer bounds of the achievable rate region of recognition systems using information theoretic arguments. While these results deepen our fundamental understanding about recognition systems, they do not provide a practical recognition system design. In \cite{OSullivan05}, a recognition system design using low-density parity-check (LDPC) matrices is proposed for independent and identically distributed (i.i.d.) binary patterns under Bernoulli noise. Yet, in general, there are few guidelines for designing recognition systems under various noise and pattern assumptions. More general coding theory results are needed.

In this paper, we establish coding theory type results for recognition system design for discrete patterns. We show that a good linear code always leads to a good recognition system design. The benefits of using linear codes are that the encoding complexity is low; there are many results on linear codes for various types of noise distributions; many linear codes have low complexity decoding algorithms, which allow one to design fast recognition algorithms. The connections established in this paper allow one to bring successful results from linear code design to recognition system design.  Under some conditions, we show that a linear encoding can outperform the inner bound of achievable rate region obtained by Westover \cite{Westover06}; see Westover andd O'Sullivan \cite{Westover06b} for more detailed analysis of achievable rates.

\section{Problem Definitions}
Three aspects of the recognition problem we consider in this paper are the \textit{environment} under which recognition takes place, the \textit{recognition system} itself, and \textit{measures of performance}. These follow the problem setting in \cite{Westover06b}.

The environment consists of six elements, denoted as
	\begin{equation}
	\label{eq:definition000a}
  \mathcal{E} = (M_{c},P_{j},P_{x},\mathcal{X},P_{y|x},\mathcal{Y}).
  \end{equation}
\(M_{c} = 2^{nR_{c}}\) is the total number of objects to be recognized; \(R_{c}\) is the pattern rate.  Each pattern is a length \(n\) sequence with each element taking values over the set \(\mathcal{X}\). Here, we consider discrete patterns that each element of a pattern takes value over \(GF(r)\). Each pattern is drawn independently from a distribution \(P_{x}\), denoted as \(x_{i}, i \in \{1,2,\cdots,M_{c}\}\). The set of all \(M_{c}\) patterns to be recognized is denoted as \(\mathcal{C}\). In the training phase, we assume that a recognition system can observe \(x_{i}\). In the testing phase, an object index \(j\) is drawn from \(\{1,2,\cdots,M_{c}\}\) based on an index distribution \(P_{j}\). The corresponding object sequence \(x_{j}\) is then presented to the recognition system with noise whose transition probability is \(P_{y|x}\), where each element of \(y\) takes values over the set \(\mathcal{Y}\). Here we assume that \(P_{j}\) is the uniform distribution. Also, the noise, denoted as \(z\), is assumed to be additive and modeled as a length \(n\) sequence over \(GF(r)\) drawn from a distribution \(P_{z}\), independent of \(X_{i}, \forall i,\) and any design of recognition systems. Hence 
	\begin{equation}
	\label{eq:definition000b}
  P_{y|x}(y|x) = P_{z}(y-x),
  \end{equation}
and the recognition system observes data
	\begin{equation}
	\label{eq:definition001}
  y=x_{j}+z, 
  \end{equation}
where the addition is under \(GF(r)\). 

A recognition system consists of a sensory compression function \(g\), a memory compression function \(f\), and a recognition algorithm \(\phi\). The sensory compression function \(g\) maps an observed \(y \in GF(r)^{n}\) to a compressed sensory data \(\sigma \in GF(r)^{nR_{s}}\), where \(R_{s}\) is defined to be the sensory compression rate. Similarly, memory compression \(f\) maps each object sequence \(x_{i} \in GF(r)^n\) to a compressed sensory data \(s_{i} \in GF(r)^{nR_{m}}\), where \(R_{m}\) is the memory compression rate. For linear encoding cases, sensory compression and memory compression are done by using matrices \(G\) of size \(nR_{s}\) by \(n\) and \(H\) of size \(nR_{m}\) by \(n\) over \(GF(r)\), such that 
	\begin{equation}
	\label{eq:definition002}
	\sigma = Gy
	\end{equation}
is the compressed sensory data and 
	\begin{equation}
	\label{eq:definition003}
	s_{i} = Hx_{i}
	\end{equation}  
is the compressed memory data of the object with index \(i\). The set of all memory data \(s_{i},i \in \{1,2,\cdots,M_{c}\}\) is denoted as \(\mathcal{S}\). We are interested in designing good recognition systems given \((R_{c},R_{m},R_{s},P_{x},P_{z})\).

The recognition algorithm \(\phi\) takes \(\mathcal{S}\) and \(\sigma\) as inputs and computes an estimate \(\hat{j}\) of the true object index. It consists of a noise estimation algorithm and an index estimation algorithm. The noise estimation algorithm is denoted as 
	\begin{equation}
	\label{eq:definition004}
	d(s_{i},\sigma):GF(r)^{nR_{m}}\times GF(r)^{nR_{y}}\rightarrow GF(r)^{n}\cup\{e\}, 
	\end{equation}
that for each object index \(i\), it computes an estimated noise under the hypothesis that the \(i\)th object is selected in the testing phase. The estimated noise of the \(i\)th object is denoted as 
	\begin{equation}
	\label{eq:definition005}
	\hat{z}_{i}=d(s_{i},\sigma). 
	\end{equation}
If the algorithm fails for the \(i\)th index, subject to some criteria of failure depending on the system design, \(d(\cdot,\cdot)\) outputs an error \(e\). After the recognition system completes noise estimation for all indexes, it proceeds to index estimation. Since an index \(j\) is chosen uniformly in the testing phase, for index estimation, the index estimation algorithm simply selects the index estimate \(\hat{j}\) to be the index associated with the largest \(P_{z}(\hat{z}_{i})\), while we define \(P_{z}(e) = 0\). This means that the recognition system rejects indexes with noise estimation error. From now on in this paper, \(j\) always denotes the true object index selected in the test phase, and \(i \in \{1,\cdots,M_{c}\}\setminus \{j\}\). 

A recognition system makes an error if \(\hat{j} \neq j\). The average probability of error of an ensemble of recognition system design is defined to be
	\begin{equation}
	\label{eq:definition006}
	P_{e}^{n} = \sum_{f,g,\mathcal{C},z}
	P(\hat{j} \neq j| \mathcal{C}, z,f,g)P_{\mathcal{C}}(\mathcal{C})P_{z}(z) P_{f,g}(f,g),
	\end{equation}
which is averaging over all realizations of \(\mathcal{C}\), \(z\), and the recognition system. Note that \(P_{f,g}(f,g)\) is specified when the ensemble of recognition system designs is defined, and 
	\begin{equation}
	\label{eq:definition007}
	P_{\mathcal{C}}(\mathcal{C}) = \prod_{x \in \mathcal{C}} P_{x}(x).
	\end{equation}
Probability of error depends on the pattern length \(n\). A three rate tuple \((R_{c},R_{m},R_{s})\) is said to be achievable in an environment \(\mathcal{E}\) if there exists a recognition system such that \(P_{e}^{n}\) goes to zero as \(n\) goes to infinity. 

\section{Truncation Encoding For i.i.d. Patterns and i.i.d. Noise}
In this section, we show that a truncation encoding outperforms the inner bounds of achievable rate region of Bernoulli \(\frac{1}{2}\) patterns under Bernoulli noise obtained in \cite{Westover06} and \cite{Westover06b}. This truncation encoding works for all \(GF(r), r \geq 2\). It is assumed that each element of a pattern sequence is independent and identically distributed (i.i.d.) drawn from a distribution \(Q_{x}\) on \(GF(r)\). Similarly, each element of the noise sequence is i.i.d. drawn from \(Q_{z}\) on \(GF(r)\). Let \(H = [I_{nR_{m}} 0]\) and \(G = [I_{nR_{s}} 0]\), where \(I_{nR_{m}}\) and \(I_{nR_{s}}\) are identity matrices of size \(nR_{m}\) and \(nR_{s}\) respectively. Thus \(s_{i}\) is the first \(nR_{m}\) elements of \(x_{i}\), and \(\sigma\) is the first \(nR_{s}\) elements of \(y=x_{j}+z\). Let \(n_{\min} = \min(nR_{m},nR_{s})\). For any length \(n\) sequence \(a\), \(a_{n_{\min}}\) denotes the sequence of the first \(n_{\min}\) elements of \(a\), and \(a_{n_{\min}^{c}}\) denotes the rest of \(a\). By definition, we know that \(s_{i,n_{min}} = x_{i,n_{min}}\) and \(\sigma_{n_{min}} = y_{n_{min}}\).

The noise estimation algorithm works as follows. For each pair of \((s_{i},\sigma)\), the algorithm checks if \((s_{i,n_{min}},\sigma_{n_{min}})\) is in the jointly typical set \(T_{n_{\min}}^{xy,\epsilon}\), where the jointly typical set \(T_{n_{\min}}^{xy}\) is defined as
	\begin{eqnarray}
	T_{n_{\min}}^{xy,\epsilon} & = & \{ (x,y) \in GF(r)^{n_{\min}} \times GF(r)^{n_{\min}} : \nonumber \\
	& &
	\left|-\frac{1}{n_{\min}}\log P(x)-H(Q_{x})\right| < \epsilon  \nonumber \\
	& & 
	\left|-\frac{1}{n_{\min}}\log P(y)-H(Q_{x}*Q_{z})\right| < \epsilon \nonumber \\
  & &
  \left|-\frac{1}{n_{\min}}\log P(x,y)-H(Q_{x})-H(Q_{z})\right| < \epsilon \nonumber \},\\
  & &
  \ \ \ 
	\end{eqnarray}
where \(Q_{x}*Q_{z}\) denotes the output distribution of a noisy channel with input distribution \(Q_{x}\) and additive noise distribution \(Q_{z}\). It proceeds if \((s_{i,n_{min}},\sigma_{n_{min}}) \in T_{n_{\min}}^{xy,\epsilon}\), otherwise it outputs an \(e\) indicating an error. The algorithm computes 
	\begin{equation}
	\label{eq:timeshare005}
	\hat{z}_{i,n_{\min}} = \sigma_{n_{\min}} - s_{i,n_{\min}} = x_{i,n_{\min}} - x_{j,n_{\min}} + z_{n_{\min}},
	\end{equation}
and then concatenates it with \(n-n_{min}\) zeros to get the estimated noise \(\hat{z}_{i}\). Finally, the systems selects the index 
	\begin{equation}
	\label{eq:estimatej001}
	\hat{j} = \arg\max_{k \in \{1,2,\cdots,M_{c}\}} P_{z}(\hat{z}_{k}) 
	= \arg\max_{k \in \{1,2,\cdots,M_{c}\}} P_{z}(\hat{z}_{k,n_{\min}}) 
	\end{equation}
as its estimated index.

\textit{Theorem 1} The probability of \(\hat{j} \neq j\) goes to zero as \(n\) goes to infinity if
	\begin{equation}
	\label{eq:theorem0101}
  R_{c} < \min(R_{m},R_{s})(H(Q_{x}*Q_{z})-H(Q_{z})-3\epsilon).
	\end{equation}

\begin{proof} There are two situations under which the truncation encoding recognition system makes an error. The first situation is when \((s_{j,n_{min}},\sigma_{n_{min}})\) is not in \(T_{n_{\min}}^{xy,\epsilon}\). The second situation is when \((s_{j,n_{min}},\sigma_{n_{min}}) \in T_{n_{\min}}^{xy,\epsilon}\) but there exists at least one other object index \(i\) such that \((s_{i,n_{min}},\sigma_{n_{min}}) \in T_{n_{\min}}^{xy,\epsilon}\) and \(P(\hat{z}_{i}) \geq P(\hat{z}_{j})\). The probability of the first situation goes to \(\epsilon\) as \(n\) goes large, and \(\epsilon\) can be chosen to be arbitrarily small because of the standard property of jointly typical set. The probability of the second situation can be bounded by the probability that there exists at least one other object \(i\) with \((s_{i,n_{min}},\sigma_{n_{min}}) \in T_{n_{\min}}^{xy,\epsilon}\). Hence the probability of the second condition is bounded by
\begin{eqnarray}
\label{equ:theorem01p}
  & & 
  \sum_{j \in {1,2,\cdots,M_{c}}} P(j) \sum_{z\in\{0,1\}^{n}} P(z) \nonumber \\
  & & \ \ \ \ \ \ \ \ \ \ \ 
  P\left( \exists i : 
  (x_{i,n_{\min}},y_{n_{min}}) \in T_{n_{\min}}^{xy,\epsilon} | z,i \right) 
  \\
  \nonumber
  & \stackrel{(a)}{=} &
  \sum_{z\in\{0,1\}^{n}}
  P(z) P ( \exists i : \\ 
  & & \ \ \ \ \ \ \ \ \ \ 
  (x_{i,n_{\min}},y_{n_{min}}) \in T_{n_{\min}}^{xy,\epsilon} | z )
  \\ 
  \nonumber
  & = &
  \sum_{z \in \{0,1\}^{n}} 
  P(z_{n_{\min}}^{c})P(z_{n_{\min}})\\
  & & \ \ \ \ \ \ \ \ \ \ \ 
  P ( \exists i :(x_{i,n_{\min}},y_{n_{min}}) \in T_{n_{\min}}^{xy,\epsilon} | z_{n_{\min}} )
  \\
  \nonumber
  & = &
  \sum_{z_{n_{\min}}^{c} \in \{0,1\}^{n-n_{\min}}} P(z_{n_{\min}}^{c}) 
  \sum_{z_{n_{\min}} \in \{0,1\}^{n_{\min}}}P(z_{n_{\min}}) \\
  & & \ \ \ \ \ \ \ \ \ \ \ 
  P ( \exists i :(x_{i,n_{\min}},y_{n_{min}}) \in T_{n_{\min}}^{xy,\epsilon} | z_{n_{\min}} )
  \\
  \nonumber
  & = &
  \sum_{z_{n_{\min}}^{c} \in \{0,1\}^{n-n_{\min}}} P(z_{n_{\min}}^{c}) \\
  & & \ \ \ \ \ \ \ \ \ \ \ 
  P ( \exists i :(x_{i,n_{\min}},y_{n_{min}}) \in T_{n_{\min}}^{xy,\epsilon})
  \\
	\label{independent}
	\nonumber
  & \stackrel{(b)}{<} &
  \sum_{z_{n_{\min}}^{c} \in \{0,1\}^{n-n_{\min}}} P(z_{n_{\min}}^{c}) \\
  & &
  \left( \sum_{i=\{2,3,\cdots,M_{c}\}} 
  P\left( (x_{i,n_{\min}},y_{n_{min}}) \in T_{n_{\min}}^{xy,\epsilon} \right) \right)
  \\
  & \stackrel{(c)}{=} &
  \left(M_{c}-1\right) 
  P\left( (x_{i,n_{\min}},y_{n_{min}}) \in T_{n_{\min}}^{xy,\epsilon} \right)
  \\
  & \stackrel{(d)}{<} &
  (1+\epsilon)2^{nR_{c}}2^{-n_{\min}(I(X;Y)-3\epsilon)}
  \\
  & \stackrel{(e)}{=} &
  (1+\epsilon)2^{-n\left(\min(R_{m},R_{s})(H(Q_{x}*Q_{z}) - H(Q_{z}))-R_{c}-3\epsilon \right)},
\end{eqnarray}
where
\begin{itemize}
\item[(a)] follows from that \(j\) is uniformly distributed and all \(x_{i}\) is independently drawn from the same distribution;
\item[(b)] follows from taking the union bound;
\item[(c)] follows from that the terms inside the parenthesis of (\ref{independent}) is independent of \(z_{n_{\min}}^{c}\);
\item[(d)] follows from the property of jointly typical set under the condition that if \(x_{i,n_{\min}}\) and \(y_{i,n_{\min}}\) are independent with the same marginals as \(P(x_{j,n_{\min}},y_{i,n_{\min}})\), then the probability that \((x_{i,n_{\min}},y_{n_{\min}}) \in T_{n_{\min}}^{xy,\epsilon} \leq 2^{-(I(X^{n_{\min}};Y^{n_{\min}})-3\epsilon)}\) \cite{Cover91}, and elements of \(x_{j,n_{\min}}\) and \(z_{n_{\min}}\) are i.i.d. hence so are elements of \(y_{n_{\min}}\) ;
\item[(e)] follows from 
	\begin{eqnarray}
	\label{eq:theorem01e}
  I(X;Y) & = & H(X+Z) - H(X+Z|X) \\
         & = & H(Q_{x}*Q_{z}) - H(Q_{z}).
	\end{eqnarray}
\end{itemize}

Thus if
	\begin{equation}
	\label{eq:timeshareAna011}
  R_{c} < \min(R_{m},R_{s})(H(Q_{x}*Q_{z}) - H(Q_{z}))-3\epsilon,
	\end{equation}
The probability of recognition error goes to zero as \(n\) goes to infinity. \end{proof}

\(\)\\
\textit{Corollary} In particular, if elements of \(x_{i}\) are drawn from i.i.d. Bernoulli \(\frac{1}{2}\), and noise is from i.i.d. Bernoulli \(q\), we have the lower bound of possible \(R_{c}\) to be
	\begin{equation}
	\label{eq:timeshareAna012}
  R_{c} < \min(R_{m},R_{s})(1-H(q))-3\epsilon,
	\end{equation}
where \(0 \leq \min(R_{m},R_{s}) \leq 1\) and \(H(q) \leq 1\). For i.i.d Bernoulli \(\frac{1}{2}\) source and any i.i.d Bernoulli \(q\) noise, this truncation encoding performs better then the ensemble of recognition system design based on LDPC matrices proposed by O'Sullivan and Lai \cite{OSullivan05}, that in \cite{OSullivan05}, it requires 
	\begin{equation}
	\label{eq:OL0501}
  R_{c} < \min(R_{m},R_{s}) - H(q) - \epsilon. 
	\end{equation}
Also notice that for \(R_{m} = R_{s} = R\), the bound (\ref{eq:timeshareAna012}) of \(R_{c}\) is above the inner bound from  \cite{Westover06} and is very close to the theoretical outer bound computed by Westover \cite{Westover06} and Westover and O'Sullivan \cite{Westover06b}. They have shown an outer bound which is a concave function of \(R\) and is very close to the straight line \(R(1-H(q))\).

Here we discuss another interesting example where the noise distribution \(Q_{z}\) is partially known.  We assume that each element of \(x_{i}\) is i.i.d. drawn from the uniform distribution over \(GF(r)\).  We assume that each element of \(z\) is i.i.d.~drawn from a distribution \(Q_{z}\), but only \(Q_{z}(0) = 1 - q\) is known (each element of \(z\) takes value \(0\) with probability \(1-q\)). We want to find the least upper bound on \(R_{c}\) among all such distributions given \(R = \min(R_{m},R_{s})\) using truncation encoding. This is a constrained optimization problem 
	\begin{equation}
	\label{eq:exp0101}
  \max_{Q_z} H(Q_{z}) \ \ \textrm{subject to} \ \ \sum_{k \in GF(r)} q_{k} = q, q_{k} \geq 0 \ 
  \forall k
	\end{equation}
where \(q_{k} = Q_{z}(k)\).  The maximum can easily be shown to be achieved for \(q_{k} = \frac{q}{r-1} \ \forall k \ne 0\). The least upper bound of \(R_{c}\) is then
	\begin{equation}
	\label{eq:exp0102}
  R\left(\log r + (1-q)\log(1-q) + q\log\left(\frac{q}{r-1}\right)\right),
	\end{equation}
where all logarithms are taken base \(2\).

Note that there are noticeable differences between recognition and lossless source coding with side information. The bits useful in recognition systems are different from bits useful for lossless source coding. Also even if a joint lossless source code is available, it might not be good for recognition. Given two correlated sequences \(x\) and \(y\), the achievable rate region of lossless source codes with side information obtained by Ahlswede and K\"orner \cite{Ahlswede75} is 
	\begin{eqnarray}
	\label{eq:Theorem0110}
  R_{x} & \geq & H(X|V), \\
  R_{y} & \geq & I(Y;V),
	\end{eqnarray}
where \(V\) is an auxiliary random variable and \(X-Y-V\) is a Markov chain. For \(x\) being Bernoulli \(\frac{1}{2}\), and \(y = x+ z\) where \(z\) is Bernoulli \(q\), \(R_{y} = 1\) and \(R_{x} = H(q)\) is an achievable rate pair to reconstruct \(x\) and hence reconstruct \(z\). However, Theorem 1 shows that it is not always necessary to reconstruct entire \(x\) or \(z\) for recognition. Also theorem 1, \cite{Westover06}, and \cite{Westover06b} all show that even if lossless coding is possible for a given recognition system with \(R_{m} = R_{x}, R_{s} = R_{y}\), it is not good for recognition if the compression rates are below the required bounds. A large sensory compression rate \(R_{s} = R_{y} = 1\) alone does not yield good performance because even if it is sufficient to reconstruct the true noise \(z\), it is not sufficient to suppress the probability that there exists another pattern which is jointly typical with a sequence matching the compressed memory and sensory data. From a linear coding point of view with \(G\) for encoding \(x\), the above argument means that the cardinality of each coset of \(G\) is too large to prevent that for all the \(2^{R_{c}}-1\) false objects, the coset \(G(x_{i}+y)\) does not contain a sequence which is jointly typical with \(x_{i}\).
\section{Linear encoding for arbitrary independent noise}
Although the truncation encoding works well for i.i.d. Bernoulli patterns under i.i.d. Bernoulli noise condition, we shall see that there exists many cases where LDPC encoding proposed in \cite{OSullivan05}, as well as several other linear codes or ensemble of linear codes, work reasonably well while no simple truncation encoding does. To see this, let us assume that  elements of patterns are i.i.d. drawn from the uniform distribution over \(GF(r)\), denoted as \(\bar{Q}_{x}\). The additive noise sequence is drawn from a distribution whose mean entropy is \(nR_{z}\) for some \(0<R_{z}<1\). Under this loose constraint which allows nonstationary noise distributions, it might not be sufficient to have good statistical properties for recognition by simply computing the first \(n_{\min}\) elements of the noise sequence. Notice that when an LDPC matrix is used for compression, the codes used are viewed as LDGM codes, which are also known to have good performance for source coding and channel coding \cite{Frias03} \cite{Zhong03}.

Under the pattern and noise assumptions stated above, if the LDPC recognition system design proposed by O'Sullivan and Lai \cite{OSullivan05} is used, the following Theorem 2 can be proved.
 
By \textit{good} ensemble for generating LDPC matrices, we mean that the ensemble and noise average block decoding error goes to zero as \(n\) goes to infinity. By \textit{good recognition system design} we mean that the ensemble and noise average recognition error goes to zero as \(n\) gets large.

\textit{Theorem 2}: If there exists an \textit{good} ensemble for generating LDPC matrices of rate \(R = \min(R_{m},R_{s})\), alone with a syndrome decoding algorithm under a noise distribution with entropy \(nR_{z}\), then there exists a \textit{good recognition system design} using the same LDPC matrix ensemble and syndrome decoding algorithm for all \(R_{c} < \min(R_{m},R_{s}) - R_{z}\). 

The proof is omitted since it follows directly from the following Theorem 3. 

\textit{Theorem 3} If there exists a good ensemble of linear codes of rate \(R = \min(R_{m},R_{s})\) and a decoding algorithm for a noise distribution with entropy \(nR_{z}\). Then for all \(R_{c} < \min(R_{m},R_{s}) - R_{z}\), there exists a good pattern recognition system design using the generator matrix of the linear block code, and the decoding algorithm as noise estimation algorithm under the same noise distribution. \\

\begin{proof}
Without loss of generality, let us assume that \(R_{m} \leq R_{s}\). Memory compression is done by using \(H\), denoting a parity check matrix generated by the linear code ensemble, such that \(s_{i} = Hx_{i}\). Sensory compression is done by a matrix \(G = [H^{T} 0]^{T}\). Let \(d(\cdot,\cdot)\) denotes the syndrome decoding associated with the linear code ensemble with typical set check. The typical set check is done by verifying if \(\hat{z}_{i}\) is in \(T_{n}^{z,\epsilon}\), where
	\begin{equation}
	\label{eq:duality0201}
  T_{n}^{z,\epsilon} = \{z:\left| \frac{1}{n}\log P(z) - R_{z} \right| < \epsilon \}
	\end{equation}
Because the probability of \(z \notin T_{n}^{z,\epsilon}\) is \(\epsilon\) which can be chosen to be arbitrarily small, and the decoding algorithm for inferring \(\hat{z_{j}}\) is good, we focus on the probability of index estimation error, similar to the proof of Theorem 1. The probability of index estimation error is less than
\begin{eqnarray}
\label{equ:duality02p}
	&   &
	\sum_{j=1}^{M_{c}}P(j) \sum_{z\in T_{n}^{z,\epsilon}}P(z)
	P\left(\exists i 
  : \hat{z}_{i} \in T_{n}^{z,\epsilon} | z,i \right) 
  \\
  & = &
  \sum_{z\in T_{n}^{z,\epsilon}}P(z)
	P\left(\exists i : \hat{z}_{i} \in T_{n}^{z,\epsilon} | z \right) 
  \\
  & = &
  \sum_{z\in T_{n}^{z,\epsilon}}P(z)
  P\left(\exists i : d(s_{i},\sigma)) \in T_{n}^{z,\epsilon} | z \right)
  \\ \nonumber
  & \stackrel{(a)}{=} &
  \sum_{z\in T_{n}^{z,\epsilon}}P(z)
  P(\exists i : \\ 
  & & \ \ \ \ \ \ \ \ \ \ \ \ 
  d(0,H(x_{i}-x_{1}+z)) \in T_{n}^{z,\epsilon} | z)
  \\
  & \stackrel{(b)}{=} &
  \sum_{z\in T_{n}^{z,\epsilon}}P(z)
  P\left(\exists i 
  : d(0,H(\tilde{x})) \in T_{n}^{z,\epsilon} \right)  
  \\
  & \stackrel{(c)}{\leq} &
  2^{nR_{c}}\sum_{z\in T_{n}^{z,\epsilon}}P(z)
  P\left(d(0,H(\tilde{x})) \in T_{n}^{z,\epsilon} \right)   
  \\
  & \stackrel{(d)}{\leq} &
  2^{nR_{c}}
  P\left(d(0,H(\tilde{x})) \in T_{n}^{z,\epsilon} \right)   
  \\
  & \leq &
  2^{nR_{c}}
  \sum_{\tilde{z}\in T_{n}^{z,\epsilon}} P(H\tilde{x}=H\tilde{z}|\tilde{z})
  \\
  & = &
  2^{nR_{c}}
  \sum_{\tilde{z}\in T_{n}^{z,\epsilon}} 2^{-nR_{m}}  
  \\
  & \stackrel{(e)}{\leq} &
  2^{-n(R_{m}-R_{z}-R_{c}-\epsilon)},
\end{eqnarray}
where
\begin{itemize}
\item[(a)] follows from the construction of \(G\) based on \(H\).
\item[(b)] is because elements of \(x_{i}\) and \(x_{1}\) both are i.i.d. from the uniform distribution over \(GF(r)\), and \(x_{i}\) and \(x_{1}\) are independent of each other and independent of \(z\), so that elements of \(x_{i}-x_{1}+z\) are also i.i.d. and uniformly distributed, denoted as \(\tilde{x}\);
\item[(c)] follows from union bound and there are totally \(2^{nR_{c}} - 1\) terms in the sum;
\item[(d)] follows from that \(\tilde{x}\) is independent of \(z\), see (b);
\item[(e)] The cardinality of \(T_{n}^{z,\epsilon}\) has upper bound \(2^{n(R_{z}+\epsilon})\).
Hence the probability of index estimation error goes to zero as \(n\) goes to infinity if
\end{itemize}
\begin{equation}
\label{duality2p}
	R_{c} < \min(R_{m},R_{s}) - R_{z} - \epsilon.
\end{equation}
\end{proof}

Note that clearly if the complexity of the decoding algorithm is \(O(f(n))\), the complexity of the recognition system per object is also \(O(f(n))\). Hence Theorem 3 not only connects good linear code design to good recognition system design, it also connects low complexity algorithms for decoding linear code to noise estimation in recognition systems. 

LDPC codes can be used for non-i.i.d. noise. For example, Eckford, Kschischang, and Pasupathy \cite{Eckford02} analyzed LDPC codes for Gilbert-Elliot Channels, which are binary symmetric channels with crossover probability depending on Markov processes, and Nicola, Alajaji, and Linder \cite{Nicola05} developed decoding algorithms for LDPC codes with a queue-based channel. Based on Theorem 3 and \cite{OSullivan05}, LDPC codes with the algorithms they developed can be used for good recognition system design for those noise models.


\section*{Acknowledgment}
This work has  been supported in part by the Office of Naval Research N000140610061.



%

\bibliography{PatternISIT07}
\bibliographystyle{IEEEtran}
\end{document}